\documentclass[12pt]{article}

\usepackage{amssymb}
\usepackage{amsmath}
\usepackage{amsthm}
\usepackage{graphicx}
\usepackage{setspace}
\usepackage{color}
\usepackage{natbib}
\usepackage{hyperref}
\usepackage{bm}
\usepackage{bbm}
\usepackage{enumitem}

\def\wh{\widehat}

\newcommand{\argmin}[1]{\underset{#1}{\operatorname{arg}
\operatorname{min}}\;}   
\def\tStar{\hbox{$y^{*}_{\tau}$}}
\def\tStarHat{\hbox{$\wh{y}^{*}_{\tau}$}}

\def\Feps{\hbox{$F_{\epsilon}$}}
\def\Fdot{\hbox{$F_{\epsilon}(\cdot)$}}
\def\FepsHat{\hbox{$\wh{F}_{\epsilon}$}}
\def\FdotHat{\hbox{$\wh{F}_{\epsilon}(\cdot)$}}
\def\g#1#2#3#4{\hbox{$G#3\cdot, \cdot \mid #1, #2 #4$}}

\def\hatg#1#2#3#4{\hbox{$\wh{G}#3\cdot, \cdot \mid #1, #2 #4$}}

\def\Gam#1#2{\hbox{$\Gamma(#1, #2)$}}
\def\hatGam#1#2{\hbox{$\wh{\Gamma}(#1, #2)$}}

\def\pione{\hbox{$\pi_1$}}
\def\pitwo{\hbox{$\pi_2$}}
\def\pitwoStar{\hbox{$\pi_{2}^*$}}
\def\piop#1#2{\hbox{$\pi_{#1}^{\hbox{\tiny #2}}$}}
\def\bpiop#1#2{\hbox{$\bpi_{#1}^{\hbox{\tiny #2}}$}}
\def\hatpitwoStar{\hbox{$\wh{\pi}_{2}^*$}}
\def\hatpiop#1#2{\hbox{$\wh{\pi}_{#1}^{\hbox{\tiny #2}}$}}
\def\bhatpiop#1#2{\hbox{$\wh{\bpi}_{#1}^{\hbox{\tiny #2}}$}}

\def\prpiS{\mbox{pr}^{\hbox{\footnotesize $\bpi$}}}
\def\prpi{\mbox{pr}^{\hbox{\footnotesize $\pi_{1}, \pi_{2}$}}}

\def\prOT#1#2{\mbox{pr}^{\hbox{\footnotesize $#1, #2$}}}

\def\bprpiop#1#2{\mbox{pr}^{\hbox{\footnotesize
$\bpi_{#1}^{\hbox{\tiny #2}}$}}} 
 
\def\bprHatpiop#1#2{\mbox{pr}^{\hbox{\footnotesize
      $\wh{\bpi}_{#1}^{\hbox{\tiny #2}}$}}} 

\def\qpi{q^{\hbox{\footnotesize $\bpi$}}}
\def\qOT#1#2{q^{\hbox{\footnotesize $#1, #2$}}}

\def\bqHatpiop#1#2{q^{\hbox{\footnotesize $\wh{\bpi}_{#1}^{\hbox{\tiny
#2}}$}}}

 \newcommand{\bma}[1]{\mbox{\boldmath $#1$}}

 \newcommand{\bA}{ {\bma{A}} }
 
 \newcommand{\bB}{ {\bma{B}} }

 \newcommand{\bH}{ {\bma{H}} }
 \newcommand{\bh}{ {\bma{h}} }

 \newcommand{\bX}{ {\bma{X}} }
 \newcommand{\bx}{ {\bma{x}} }

 \newcommand{\bbeta}{ {\bma{\beta}} }
 
 \newcommand{\bpi}{ {\bma{\pi}} }
 
 \newcommand{\btheta}{ {\bma{\theta}} }

 \newcommand{\bgamma}{ {\bma{\gamma}} }

\newcommand{\T}{\intercal}

\newenvironment{frcseries}{\fontfamily{frc}\selectfont}{}

\def\independenT#1#2{\mathrel{\rlap{$#1#2$}\mkern3mu{#1#2}}}
\newcommand\independent{\protect\mathpalette{\protect\independenT}{\perp}}

\newtheorem{thm}{Theorem}[section]
\newtheorem{lem}[thm]{Lemma}

\theoremstyle{definition}
\newtheorem{rmrk}[thm]{Remark}

\usepackage{JASA_manu}

\begin{document}

\title{Interactive Q-learning for Probabilities and Quantiles}

\author{Kristin A. Linn$^{1}$, Eric B. Laber$^{2}$, Leonard A. Stefanski$^{2}$ \\
$^{1}$Department of Biostatistics and Epidemiology \\  
University of Pennsylvania, Philadelphia, PA 19104 \\  
$^{2}$Department of Statistics \\  North Carolina State University, Raleigh,
NC 27695  \\  
email: \texttt{klinn@upenn.edu} }

\maketitle

\newpage

\mbox{}
\vspace*{2in}
\begin{center}
\textbf{Author's Footnote:}
\end{center}
Kristin A. Linn is Postdoctoral Fellow, Department of Biostatistics
and Epidemiology, University of Pennsylvania.  Mailing address: 
University of Pennsylvania School of Medicine,
Center for Clinical Epidemiology and Biostatistics (CCEB),
513 Blockley Hall, 
423 Guardian Drive, Philadelphia, PA 19104
(email: klinn@upenn.edu); Eric B. Laber is Assistant Professor,
Department of Statistics, North Carolina State University;  Leonard
A. Stefanski is Drexel Professor, 
Department of Statistics, North Carolina State University.  Kristin
Linn acknowledges former support from NIH traning grant 
5T32GM081057 while preparing this manuscript. Eric 
Laber 
acknowledges support from NIH grant P01 CA142538 and DNR grant
PR-W-F14AF00171. Leonard Stefanski 
acknowledges support from NIH grants R01 CA085848 and P01 CA142538 and
NSF grant DMS-0906421. 

\newpage
\begin{center}
\textbf{Abstract}
\end{center}
 A dynamic treatment regime is a sequence of decision rules 
each of which recommends treatment based on
features of patient medical history such as past treatments and
outcomes. Existing methods for estimating optimal dynamic treatment 
regimes from data optimize the mean of a 
response variable.  However, the mean may not always be the most
appropriate summary of 
performance. We derive estimators of decision rules for optimizing
probabilities and quantiles computed with respect to the response
distribution for 
two-stage, binary treatment settings. This enables
estimation of dynamic treatment regimes that optimize the cumulative 
distribution function of the response at a prespecified point or a
prespecified quantile of the response distribution such as the
median. The proposed methods perform 
favorably in simulation experiments. We illustrate our approach with
data from a sequentially randomized trial where the primary  
outcome is remission of depression symptoms. 

\vspace*{.3in}


\newpage

\section{Introduction}
\label{sec:Introduction}
\label{sec:One}

A dynamic treatment regime operationalizes clinical decision
making as a series of decision rules that dictate treatment over
time. 
These rules account for accrued patient medical history,
including past treatments and outcomes.  Each rule maps current
patient characteristics to a recommended treatment, hence
personalizing treatment.  Typically, a dynamic
treatment regime is
estimated from data with the goal of optimizing the expected value of 
a clinical outcome, and the resulting regime is referred to
as the estimated optimal regime.

Direct-search, also
known as policy-search or value-search, is one approach to 
estimating an optimal dynamic treatment regime.  Direct search
estimators require a pre-specified class of dynamic treatment regimes
and an estimator of the marginal mean outcome under any regime
in the pre-specified class.  
The maximizer of the estimated marginal mean outcome
over the class of regimes is taken as the estimator of the
optimal dynamic treatment regime.   Marginal structural
models (MSMs) are one type of direct-search estimators 
\citep{rob:00, van+Pet:05, van:06, van+Pet:07,
    Bem+van:08, Rob+etal:08, Ore+etal:10, Pet+etal:14}. 
MSMs are best-suited to problems with a small class of
potential regimes. MSMs may also be advantageous in practice because
optimizing over a small class of pre-specified 
regimes provides a simpler, and often more interpretable, regime
than other approaches.  Another class of direct-search
estimators casts  
the marginal mean outcome as a weighted missclassification
rate and applies either discrete-optimization or classification 
algorithms to optimize a plugin estimator of the marginal
mean outcome \citep{Zhao+etal:12, Zha+etal:12, Zha+etal:12b, Zha+etal:13,
Zha+Zen+etal:15}.  

Regression-based or indirect estimators comprise a second 
class of estimators of an optimal dynamic treatment regime.
Regression-based estimators require a model for some portion
of the conditional distribution of the outcome given treatments
and covariate information.  Examples of regression-based
 estimators include $Q$-learning \citep{Wat:89,   
  Wat+Day:92, Mur:05b}, regularized $Q$-learning \citep{Moo+etal:10, 
  Cha+etal:10, Son+etal:11, Gol+etal:12}, Interactive $Q$-learning
\citep[][]{Lab+etal:13}, $g$-estimation in structural nested 
mean models \citep{Rob:04}, 
$A$-learning \citep{Mur:03}, and regret-regression
\citep{Hen+etal:10}.  
Regression-based approaches often target the globally optimal regime
rather than restricting attention to a small, pre-specified
class. They can be also be useful in exploratory contexts to discover
new treatment strategies for further evaluation in later trials.

Direct-search and regression-based
estimators have been extended to handle survival outcomes
  \citep{Gol+Kos:12,Hua+Nin:12, Hua+etal:14},
high-dimensional data \citep{McK+Qia:13}, 
missing data \citep{Sho+etal:14}, and 
multiple outcomes \citep{Lab+Liz+etal:14, kristinBook,
  luedtke2015optimal}.

Despite many estimation methods, none are designed 
to handle functionals of the response distribution other than the
mean, such as probabilities or quantiles.  The median response is
often of interest in studies where the outcome follows a skewed
distribution, such as the total time a women spends in second stage
labor \citep{Zha+Che:12}.
Using the potential
outcomes framework \citep{Rub:74, Ros+Rub:83}, \citet{Zha+Che:12}  
develop methods for estimating quantiles of the 
potential outcomes from observational data.  However, they focus on
comparing treatments at a single intervention time point rather than
estimation of an optimal dynamic treatment regime. Structural nested
distribution models (SNDMs) estimated using g-estimation facilitate
estimation of point treatment effects on the cumulative distribution
function of the outcome \citep{rob:00, van+jof:14}. Thus far, SNDMs
have not been extended to 
estimate a regime that maximizes a threshold exceedence probability or
quantile.

$Q$-learning and its variants are often useful when targeting the
globally optimal regime because they provide relatively interpretable
decision rules that are based on (typically linear) 
regression models. However, the
$Q$-learning algorithm is an approximate dynamic programming   
procedure that requires modeling nonsmooth, nonmonotone
transformations of data. This leads to nonregular estimators 
and complicates the search for models that fit the 
data well \citep{Rob:04,
  Cha+etal:10, Lab+etal:10, Son+etal:11}. Interactive $Q$-learning
(IQ-learning), developed 
for the two-stage binary treatment setting, requires modeling only 
smooth, monotone transformations of the data, thereby reducing
problems of model 
misspecification and nonregular inference \citep{Lab+etal:13}.  
We extend the IQ-learning framework to
optimize functionals of the outcome distribution other than the
expected value. In particular, we optimize threshold-exceedance
probabilities and quantiles of the response distribution.  
Furthermore, because this extension of 
IQ-learning provides an estimator of a 
threshold-exceedance probability or quantile of the response
distribution under any postulated dynamic treatment regime, 
it can be used to construct direct-search estimators.  

Threshold-exceedance probabilities are relevant in clinical
applications where the primary objective is remission or a specific
target for symptom reduction. For example, consider a population of
obese patients enrolled in a study to determine the effects of several
treatment options for weight loss.  The treatments of interest may
include combinations of drugs, exercise programs, counseling, and meal
plans \citep[][]{Ber+etal:10}.  Our method can be used
to maximize the probability that patients achieve a weight below some
prespecified, patient-specific threshold at the conclusion of the
study.  Optimization of threshold-exceedance probabilities can
be framed as a special case optimizing the mean of a binary
outcome; however, we show that for a large class of
simple generative models, $Q$-learning for binary 
data \citep{Cha+Moo:13} applied to threshold-exceedance 
indicators yields an estimator that is constant
across all threshold values whereas the true optimal regime
changes drastically across threshold values.  


With adjustments to the method of maximizing probabilities, we also
derive optimal decision rules for maximizing quantiles of the response
distribution.  Both frameworks can be used to study the {\em entire
  distribution} of the outcome under an optimal dynamic treatment
regime; thus, investigators can examine how the optimal regime changes
as the target probability or quantile is varied. In addition, the
quantile framework provides an analog of quantile regression in the
dynamic treatment regime setting for constructing robust estimators;
for example, it enables optimization of the median response.


\section{Generalized Interactive $Q$-learning}
\label{subsec:Sec2Intro}
We first characterize the optimal regime for a probability
and quantile using potential outcomes \citep[][]{Rub:74} and two
treatment time-points. We assume that the observed data,
$\mathcal{D} = \lbrace (\bX_{1i}, A_{1i}, \bX_{2i}, A_{2i},
Y_{i})\rbrace_{i=1}^{n}$, comprise $n$ independent,
identically distributed, time-ordered trajectories; 
one per patient. 
Let $(\bX_1, A_1, \bX_2, A_2, \allowbreak
Y)$ denote a generic observation where:
$\bX_1 \in \mathbb{R}^{p_1}$ is
baseline  
covariate information collected prior to the first treatment; 
$A_1 \in
\lbrace -1, 
1\rbrace$ is the first treatment; 
$\bX_2\in\mathbb{R}^{p_2}$ is interim
covariate information collected  
during the course of the first treatment but prior 
the second treatment; 
$A_2 \in \lbrace -1, 1\rbrace$ is the second
treatment; and $Y\in \mathbb{R}$ is an outcome measured at the
conclusion of stage two, coded so that larger is better.  Define
$\bH_{1} = \bX_{1}$ and  
$\bH_{2} = (\bH_{1}^{\T}, A_{1}, \bX_{2}^{\T})^{\T}$ so
that $\bH_{t}$ is the information available to a decision
maker at time $t$.
A regime, $\bpi = (\pi_{1}, \pi_{2})$, is a pair of decision rules
where $\pi_{t}: \mbox{dom}(\bH_{t}) \mapsto \mbox{dom}(A_{t})$,
such that a patient presenting
with $\bH_{t}=\bh_{t}$ at time $t$ is recommended 
treatment $\pi_{t}(\bh_{t})$.

Let $\bH_2^*(a_1)$ be the potential second-stage history under
treatment $a_1$ and $Y^*(a_1, a_2)$ the potential outcome under
treatment sequence $(a_1, a_2)$.  Define the set of all potential
outcomes $W = \left\lbrace \bH_2^{*}(a_1), Y^*(a_1, a_2)\,:\, (a_1,
  a_2)\in \lbrace -1,1\rbrace^{2} \right\rbrace$.  Throughout we
assume: (C1) consistency, so that $Y = Y^*(A_1, A_2)$; (C2) sequential
ignorability \citep[][]{Rob:04}, i.e., $A_t \independent W \mid \bH_t$ for
$t=1,2$; and (C3) positivity, so that there exists $\epsilon > 0$ for
which 
$\epsilon < \mbox{pr}(A_t = a_t|\bH_t) < 1-\epsilon$ with probability one for
all $a_t,\,t=1,2$.  Assumptions (C2)-(C3) hold by design when data are
collected using a sequential multiple assignment randomized trial
\citep[SMART,][]{Lav+Daw:00, Lav+Daw:04, Mur:05a}.  In observational
studies, these assumptions are not testable.   
We assume that
data are collected using a two-stage, binary treatment SMART.
This set-up facilitates a focused discussion of the proposed 
methods and is also useful in practice, as data in many
sequentially randomized trials have this structure \citep{psu:12,
  eblSmart}.  However, the next two results, which are
proved in the supplemental material, demonstrate that the 
proposed methodology can be extended to observational data 
and studies with more than two treatments.  

For any $\bpi$ define $Y^*(\bpi) = \sum_{(a_1,a_2)}Y^*(a_1, a_2)
\mathbbm{1}_{\pi_1(\hbox{\scriptsize $\bH_1$})=
  a_1}\mathbbm{1}_{\pi_2\{\hbox{\scriptsize 
    $\bH_2^*(a_1)$} \}=a_2}$ to be the
potential outcome under $\bpi$.  Define the function
$R(y;\bx_1, a_1, \bx_2, a_2) = \mbox{pr} (Y\ge y|\bX_1=\bx_1, A_1=a_1,
\bX_2=\bx_2, A_2=a_2)$ for any $y\in\mathbb{R}$.  The following 
result expresses the survival function
of $Y^*(\bpi)$ in terms of the underlying generative model and is
proved in the supplemental material.
\begin{lem}
Assume (C1)-(C3) and let $y\in\mathbb{R}$.  Then, for any 
$\bpi$
\begin{equation*}
\mbox{pr}\left\lbrace
Y^*(\bpi)\ge y
\right\rbrace = \mathbb{E}\left[
\sum_{a_1}\mathbbm{1}_{\pi_1(\hbox{\scriptsize
    $\bH_1$})=a_1}\mathbb{E}\left\lbrace 
\sum_{a_2}\mathbbm{1}_{\pi_2(\hbox{\scriptsize
    $\bH_2$})=a_2}R(y;\bX_1, a_1, \bX_2, a_2) 
\big| \bX_1, A_1=a_1
\right\rbrace
\right].
\end{equation*}
\end{lem}\noindent
This result, which is essentially the $g$-computation formula
\citep[][]{robins1986new}, shows that 
$\mbox{pr}\left\lbrace Y^*(\bpi)\ge y\right\rbrace$ is maximized by
$\bpi^{y}=(\pi_{1}^{y}, \pi_{2}^{y})$ where $\pi_2^{y}(\bh_2) = 
\arg\max_{a_2}R(y;\bx_1, a_1, \bx_2, a_2)$ and
$\pi_1^{y}(\bh_1) = \arg\max_{a_1}\mathbb{E}
\left\lbrace \sum_{a_2}\mathbbm{1}_{\pi_2^{y}(\hbox{\scriptsize
      $\bH_2$})=a_2}R(y; \bx_1, a_1,\bX_2, a_2)  
\right\rbrace$.  This lemma can also be used to characterize
the regime that optimizes a quantile.  Define
$\Psi = \left\lbrace \bpi^{y}\,:\,y\in\mathbb{R}\right\rbrace$.
The following result is proved in the supplemental material.
\begin{thm}\label{quantileThm}
Assume (C1)-(C3) and that the map $y \mapsto R(y;\bx_1, a_1, 
\bx_2, a_2)$ from $\mathbb{R}$ into $(0,1)$
is surjective for all $\bx_1, a_1, \bx_2$, and $a_2$.
Let $\zeta > 0$ and $\tau\in(0,1)$ be arbitrary. Then, there exists 
$\widetilde{\bpi} \in \Psi$ so that
$\inf\left\lbrace y \,:\, \mbox{pr}\left\lbrace Y^*(\widetilde{\bpi})
    \le y\right\rbrace 
\ge \tau \right\rbrace \ge 
\sup_{\pi}\inf\left\lbrace y \,:\, \mbox{pr}\left\lbrace Y^*(\bpi) \le
    y\right\rbrace 
\ge \tau \right\rbrace - \zeta$.  
\end{thm}

\subsection{Threshold Interactive $Q$-learning}
\label{subsec:TIQLearning}
Our estimators are developed under the following setup.  Because
$A_{2}$ is binary, there exist functions $m$ and $c$ such that $E(Y
\mid A_{2}, \bH_{2}) = m(\bH_{2}) + A_{2}c(\bH_{2})$.  We assume that
$Y = E(Y \mid A_{2}, \bH_{2}) + \epsilon$, where 
$\epsilon \sim (0, \sigma^2)$ independently of $(A_2, \bH_2)$.  
In the supplemental
material we describe extensions to: (i) non-additive error structures;
and (ii) heteroskedastic error, i.e., $\mathrm{Var}(\epsilon\mid
\bH_2=\bh_2, A_2=a_2) = \sigma^2(\bh_2, a_2)$ for unknown function
$\sigma$.

Let $\mbox{pr}^{\hbox{\scriptsize $\bpi$}}(Y > \lambda)$, equivalently
$\prpi(Y > \lambda)$, denote 
the probability that the outcome $Y$ 
is greater than a predefined threshold $\lambda$ under treatment
assignment dictated by the regime $\bpi = (\pi_{1}, \pi_{2})$.
Threshold Interactive $Q$-learning (TIQ-learning) maximizes
$\mbox{pr}^{\hbox{\scriptsize $\bpi$}}\{Y 
> \lambda(\bH_{t}) \mid \bH_1 \}$ for all $\bH_1$ with respect to
$\bpi$, where $\lambda(\bH_{t})$ is a threshold
that depends on $\bH_{t}$, $t=1,2$.  Here, we assume $\lambda(\bH_{t})
\equiv \lambda$; patient-specific thresholds are discussed in the
supplemental material.  As $\mbox{pr}^{\hbox{\scriptsize $\bpi$}}(Y >
\lambda\mid \bH_1) = 
E^{\hbox{\scriptsize $\bpi$}}\{ \mathbbm{1}_{Y
  >\lambda} \mid \bH_1 \}$, one approach to estimating an optimal
regime is to use discrete $Q$-learning with the outcome $\mathbbm{1}_{Y >
  \lambda}$. However, we show analytically in Remark \ref{discreteQ},
and empirically in Section \ref{sec:montecarloresults}, that in many
cases, discrete $Q$-learning is equivalent to $Q$-learning with
outcome $Y$ and is therefore insensitive to the threshold $\lambda$.


Define $F_{\hbox{\scriptsize $\bH_{1}$}}(\cdot)$ to be the 
distribution of 
$\bH_{1}$; $F_{\hbox{\scriptsize $\bH_{2} \mid \bH_{1},
    A_{1}$}}(\cdot \mid \bh_{1}, 
a_{1})$ to be the conditional distribution of
$\bH_{2}$ given $\bH_{1}=\bh_{1}$ and $A_{1}=a_{1}$; 
$\Fdot$ to be the distribution of $\epsilon$; and 
$\bH_{2}^{\hbox{\footnotesize $\pi_{1}$}(\hbox{\scriptsize $\bH_{1}$})} 
= \{ \bH_{1}^{\T}, 
\pi_{1}(\bH_{1}), \bX_{2}^{\T}\}^{\T}$. 
Let  $J^{\hbox{\footnotesize $\pi_{1}, \pi_{2}$}}(\bh_{1}, \bh_{2}, y)
  = \Feps \{y - m(\bh_{2}^{\hbox{\footnotesize
      $\pi_{1}$}(\hbox{\scriptsize $\bh_{1}$})})  -  
\pi_{2}(\bh_{2}^{\hbox{\footnotesize $\pi_{1}$}(\hbox{\scriptsize
    $\bh_{1}$})})c(\bh_{2}^{\hbox{\footnotesize
    $\pi_{1}$}(\hbox{\scriptsize $\bh_{1}$})})\}$,   
then
\begin{equation}
\label{piProb}
\prOT{\pione}{\pitwo}(Y \leq y) = \int \int
J^{\hbox{\footnotesize $\pi_{1}, 
  \pi_{2}$}}(\bh_{1}, \bh_{2}, y) dF_{\hbox{\scriptsize
    $\bH_{2}\mid \bH_{1}, A_{1}$}}\{\bh_{2} \mid 
\bh_{1}, \pi_{1}(\bh_{1})\}  dF_{\hbox{\scriptsize $\bH_{1}$}}(\bh_{1}),
\end{equation} 
is the expected
value of $J^{\hbox{\footnotesize $\pi_{1}, \pi_{2}$}}(\bH_{1}, \bH_{2},
  y)$. 

Let
$\pitwoStar(\bh_2) = \mbox{sgn}\{
c(\bh_2)\}$, where $\mbox{sgn}(x) = \mathbbm{1}_{x
  \geq 0} - \mathbbm{1}_{x < 0}$. Then, 
$J^{\hbox{\footnotesize $\pi_{1}$},
  \hbox{\footnotesize $\pitwoStar$}}(\bh_{1}, \bh_{2}, y) = \Feps \{y - 
m(\bh_{2}^{\hbox{\footnotesize
    $\pi_{1}$}(\hbox{\scriptsize $\bh_{1}$})})  -
|c(\bh_{2}^{\hbox{\footnotesize
    $\pi_{1}$}(\hbox{\scriptsize $\bh_{1}$})})|\}$ and
$\pitwo(\bh_{2}^{\hbox{\footnotesize
    $\pi_{1}$}(\hbox{\scriptsize
    $\bh_{1}$})})c(\bh_{2}^{\hbox{\footnotesize 
    $\pi_{1}$}(\hbox{\scriptsize $\bh_{1}$})}) 
\leq |  c(\bh_{2}^{\hbox{\footnotesize
    $\pi_{1}$}(\hbox{\scriptsize $\bh_{1}$})})|$ for all
$\bh_{2}^{\hbox{\footnotesize
    $\pi_{1}$}(\hbox{\scriptsize $\bh_{1}$})}$, 
implies
\begin{eqnarray}
\label{opt2}
\prOT{\pione}{\pitwo}(Y \leq y) &\geq& \int \int
J^{\hbox{\footnotesize $\pi_{1}$},  \hbox{\footnotesize
    $\pitwoStar$}}(\bh_{1}, \bh_{2}, y) 
dF_{\hbox{\scriptsize $\bH_{2} \mid 
  \bH_{1}, A_{1}$}}\{\bh_{2} \mid 
\bh_{1}, \pi_{1}(\bh_{1})\}  dF_{\hbox{\scriptsize
    $\bH_{1}$}}(\bh_{1}),
\end{eqnarray}
where the
right-hand side of \eqref{opt2} is
$\prOT{\pione}{\pitwoStar}(Y \leq y)$. Let $G(\cdot, \cdot \mid \bh_1,  
a_1)$ denote the joint conditional 
distribution of $m(\bH_{2})$ and $c(\bH_{2})$ 
given $\bH_{1}=\bh_{1}$ and $A_{1}=a_{1}$, then 
$\prOT{\pione}{\pitwoStar}(Y \leq
y) = E \left( I \left[ y,
    \Fdot, \g{\bH_1}{\pi_{1}(\bH_1)}{\{}{\}} \right] \right)$,
where


\begin{equation}
\label{It}
I\{y, \Fdot, \g{\bh_1}{a_1}{(}{)} \} = \int
\Feps(y - u - |v|)dG(u, v \mid \bh_1, a_1).
\end{equation}
The $\lambda$-optimal regime $\bpiop{\lambda}{TIQ} = \{ \piop{1, 
  \lambda}{TIQ}, \piop{2, \lambda}{TIQ} \}$ 
satisfies $\bprpiop{\lambda}{TIQ} (Y > \lambda) \geq \prpiS   
(Y > \lambda)$ for all $\bpi$. 
That is, the
distribution of $Y$ induced by regime $\bpiop{\lambda}{TIQ}$ has at 
least as  much
mass above $\lambda$ as 
the distribution of $Y$ induced by any other regime. It
follows from the lower bound on $\prOT{\pione}{\pitwo}(Y \leq y )$ 
displayed in  \eqref{opt2} that
$\piop{2,\lambda}{TIQ}(\bh_{2}) 
=\pitwoStar(\bh_{2}) = \mbox{sgn}\{c(\bh_{2})\} $ for all 
$\bh_{2}$, independent of $\lambda$ and
$\piop{1,\lambda}{TIQ}$. Henceforth, we  denote
$\piop{2,\lambda}{TIQ}$ by $\pitwoStar$. The relationship
\begin{eqnarray}
\label{argA1}
\prOT{\pione}{\pitwoStar} (Y > \lambda) &=& 1 - E
\left( I \left[ \lambda, \Fdot, \g{\bH_1}{\pi_{1}(\bH_1)}{\{}{\}}
  \right] 
\right) \nonumber \\ 
&\leq& 1 - E\left[ \min_{a_1} I\{ \lambda, \Fdot, \g{\bH_1}{a_1}{(}{)} 
  \} \right], 
\end{eqnarray}
shows that the $\lambda$-optimal first-stage rule is 
$\piop{1, \lambda}{TIQ}(\bh_1) = \argmin{a_1} I\{ \lambda,
\Fdot, \g{\bh_1}{a_1}{(}{)} \}$. Inequality \eqref{argA1} 
holds because $I\{\lambda, \Fdot,  \g{\bH_1}{a_1}{(}{)} \}$ is
minimized over $a_{1}$ for all $\bH_{1}$. It will be useful
later on to write $\piop{1, \lambda}{TIQ}(\bh_1) = \mbox{sgn}\left\{ d
  (\bh_{1}, \lambda) \right\}$ where
\begin{equation}
\label{dFn}
 d(\bh_{1}, \lambda) = I\{ \lambda, 
\Fdot, \g{\bh_1}{-1}{(}{)} \}  - I\{ \lambda,
\Fdot, \g{\bh_1}{1}{(}{)} \}.
\end{equation}

Below, we describe the
general form of the TIQ-learning algorithm that  
can be  used to
estimate the $\lambda$-optimal regime. The exact   
algorithm depends on the choice of estimators for $m(\bH_{2})$,
$c(\bH_{2})$, \Fdot\ and
\g{\bh_1}{a_1}{(}{)}. For example, one might posit parametric models,
$m(\bH_{2}; \bbeta_{2,0})$ and $c(\bH_{2}; \bbeta_{2,1})$, for
$m(\bH_{2})$ and $c(\bH_{2})$ and estimate the parameters in the model
$Y = m(\bH_{2}; \bbeta_{2,0}) + c(\bH_{2}; \bbeta_{2,1}) + \epsilon$ using least
squares. Alternatively, these terms could be estimated 
nonparametrically.
We discuss possible estimators for \Fdot\ and \g{\bh_{1}}{a_{1}}{(}{)} in
Sections \ref{subsec:F} and \ref{subsec:g}. In practice, the  
choice of 
estimators should be informed by the observed data. Define
$\wh{d}(\bh_{1}, \lambda) = I\{\lambda, \FepsHat(\cdot), \wh{G}(\cdot,  
\cdot \mid \bh_{1}, -1)\} - I\{\lambda, \FepsHat(\cdot), \wh{G}(\cdot, 
\cdot \mid \bh_{1}, 1)\}$.

\vspace{3mm}

\underline{\bf TIQ-learning algorithm:} 

\vspace{-2mm}

\begin{enumerate}[leftmargin=.8in]
\item[TIQ.1] Estimate $m 
(\bH_{2})$ and $c (\bH_{2})$, and denote the resulting estimates by
$\wh{m}(\bH_{2})$ and $\wh{c}(\bH_{2})$. Given $\bh_{2}$, estimate
$\pitwoStar$ 
using the plug-in estimator 
$\hatpitwoStar(\bh_{2}) = \mbox{sgn}\{\wh{c}(\bh_{2})\}$.

\item[TIQ.2] Estimate $\Fdot$, the cumulative distribution function of
$\epsilon$, using the residuals $\hat{e}^{Y} = Y - \wh{m}(\bH_{2}) -
A_{2}\wh{c}(\bH_{2})$ from TIQ.1. Let
$\FdotHat$ denote this estimator.

\item[TIQ.3] Estimate $\g{\bh_1}{a_1}{(}{)}$, the joint
conditional 
distribution of $m(\bH_{2})$ and $c(\bH_{2})$ given $\bH_{1}=\bh_{1}$ 
and 
$A_{1}=a_{1}$. Let $\hatg{\bh_{1}}{a_{1}}{(}{)}$ denote  this
estimator. 

\item[TIQ.4] Given $\bh_{1}$, estimate
$\piop{1,\lambda}{TIQ}$ 
using the plug-in estimator $\hatpiop{1,
  \lambda}{TIQ}(\bh_{1})  = \mbox{sgn}\{\wh{d}(\bh_{1}, \lambda)\}$. 
\end{enumerate}

The TIQ-learning algorithm involves modeling $m(\cdot)$, $c(\cdot)$,
the distribution function $\Fdot$, and the bivariate conditional
density $\g{\bh_1}{a_1}{(}{)}$.  This is more modeling than some
mean-targeting algorithms.  For example, $Q$-learning involves
modeling $m$ and $c$, and the conditional mean of
$m(\bH_{2}) + |c(\bH_{2})|$ given $\bH_1$, $\bA_1$, and $IQ$-learning
involves modeling $m$, $c$, the conditional mean of $m(\bH_2)$, and
the conditional density of $c(\bH_{2})$ \citep{Lab+etal:13}.  We
discuss models for the components of the TIQ-learning algorithm in
Sections \ref{subsec:F} and \ref{subsec:g}.


\begin{rmrk}\label{parNonPar}
Standard trade-offs between parametric and nonparametric estimation
apply to all terms in the TIQ-learning algorithm. In practice, the
choice of estimators will likely depend on sample size and the
scientific goals of the study.  If the goal is to estimate
a regime for immediate decision support in the clinic, then
the marginal mean outcome of the estimated regime is
of highest priority.  Given sufficient data, it may be
desirable to use nonparametric estimators with TIQ-learning
in this context. However, if the goal is to inform future research
and generate hypotheses for further investigation, then 
factors like parsimony, interpretability, and the ability 
to identify and test for factors associated with heterogeneous
treatment response may be most important.  In this context,
parametric models may be preferred for some or all of the
components of the TIQ-learning algorithm.   
\end{rmrk}


\begin{rmrk}\label{diffOpt}
Let $\pi_{1}^{\hbox{\scriptsize M}}$ denote the first-stage decision
rule of an 
optimal regime for the mean of $Y$.  Then, assuming the set-up of
Section 
\ref{subsec:Sec2Intro}, it can be shown
\begin{eqnarray*}
\pi_{1}^{\hbox{\scriptsize M}} (\bh_{1}) = \arg\min_{a_{1}}\int (-u -
|v|)dG(u, v \mid \bh_{1}, a_{1}) = \arg\min_{a_{1}} \int (\lambda -u -
|v|)dG(u, v \mid \bh_{1}, a_{1}), 
\end{eqnarray*}
whereas $\piop{1, \lambda}{TIQ}(\bh_{1}) = \arg\min_{a_{1}}\int
F_{\epsilon}(\lambda - u - |v|) dG(u, v \mid \bh_{1}, a_{1})$. 
If $F_{\epsilon}(\cdot)$ is approximately linear where 
the conditional distribution of $\lambda - m(\bH_{2}) - |c(\bH_{2})|$
given $\bH_{1} = \bh_{1}$ and $A_{1} = a_{1}$ is
concentrated, $\pi_{1}^{\hbox{\scriptsize M}} (\bh_{1})$ and $\piop{1,
  \lambda}{TIQ}(\bh_{1})$ will likely agree.
Thus, the difference between the mean
optimal and TIQ-learning optimal regimes can be compared
empirically by computing 
$\arg\min_{a_{1}}\int (-u -
|v|)d\wh{G}(u, v \mid \bh_{1i}, a_{1})$, $\arg\min_{a_{1}}\int
\FepsHat(\lambda -u - 
|v|)d\wh{G}(u, v \mid \bh_{1i}, a_{1})$,
for each first-stage patient history $\bh_{1i}$, $i=1, \dots, n$, and
examining where these rules differ.
\end{rmrk}

\begin{rmrk}\label{discreteQ}
  One approach to estimating an optimal decision rule
  for threshold-exceedance probabilities is discrete
  $Q$-learning \citep{Cha+Moo:13}. However, under certain
  generative models, the
  estimand in discrete $Q$-learning is the same as the
  estimand in $Q$-learning using the continuous outcome $Y$ and is
  therefore 
  independent of the threshold $\lambda$. 
  Hence, discrete $Q$-learning need 
  not be consistent when the optimal decision rule
  depends on the choice of threshold.  
  Discrete $Q$-learning assumes that
    $L\{\mbox{pr}(\mathbbm{1}_{Y > \lambda} = 1 \mid \bH_{2},
    A_{2})\} = m^{*}(\bH_{2}) + A_{2}c^{*}(\bH_{2})$,
where $L$ is a monotone increasing link function, 
and $m^*$ and $c^*$ are (unknown) functions of $\bH_2$. 
The estimated optimal second-stage decision rule
is $\widetilde{\pi}_{2}^{DQ}(\bh_2) = \mathrm{sign}\left\lbrace c^*(\bh_2)
\right\rbrace$.  
Defining $\widetilde{Y} = m^*(\bH_2) + |c^*(\bH_2)|$,
the first-stage optimal decision rule in
discrete $Q$-learning is then defined as 
$\widetilde{\pi}_{1}^{DQ}(\bh_1) = \arg\max_{a_1}
\mathbb{E}\left\lbrace \widetilde{Y}
\mid \bH_1=\bh_1,A_1=a_1
\right\rbrace$.   Suppose that 
$Y = m^*(\bH_2) + A_2c^*(\bH_2) + \epsilon$, so that
discrete $Q$-learning is correctly specified at the 
second stage if $L(u) = \lambda - F_{\epsilon}^{-1}(1-u)$, where 
$F_{\epsilon}$ is the cumulative distribution function of
$\epsilon$.  Then, under
this specification, $\widetilde{Y} = \max_{a_{2}}\mathbb{E}\left(Y|
\bH_2, a_2\right)$ which: (i) does not depend on 
 $\lambda$; and (ii) makes the estimands in discrete
$Q$-learning identical to the estimands in $Q$-learning
targeted at optimizing the mean of $Y$.  Thus, under
such generative models, we expect discrete $Q$-learning
and $Q$-learning to perform similarly across all values of
 $\lambda$.  We demonstrate this using simulation
experiments in Section \ref{sec:montecarloresults}.
\end{rmrk}


\subsection{Quantile Interactive $Q$-learning}
\label{subsec:QIQlearning}

Under some generative models, assigning treatment according to
a mean-optimal regime leads to higher average outcomes at
the expense of higher variability, negatively
affecting patients with outcomes in the lower quantiles of the induced
distribution of 
$Y$. We demonstrate this using simulated examples in Section
\ref{sec:montecarloresults}.  
Define the $\tau^{\hbox{\scriptsize th}}$ quantile of the distribution
of $Y$ induced by regime
$\bpi$ as
$\qpi(\tau) = \inf \{ y: \prpi(Y \leq y) \geq
\tau \}$. The goal of Quantile Interactive $Q$-learning (QIQ-learning)
is to  estimate a pair of decision
  rules, $\bpiop{\tau}{QIQ} =
\{\piop{1, \tau}{QIQ}, \piop{2, \tau}{QIQ}\}$, that
maximize $\qpi (\tau)$ over $\bpi$ for a fixed, prespecified
$\tau$. QIQ-learning  is
similar to TIQ-learning, but the
optimal first-stage rule is complicated by the
inversion of the distribution function to obtain
quantiles of $Y$ under a given regime. Under the model assumptions of 
Section \ref{subsec:Sec2Intro}, the QIQ-learning
second-stage  
optimal decision is 
$\piop{2, \tau}{QIQ}(\bh_{2}) =  \pitwoStar(\bh_2) =
\mbox{sgn}\{c(\bh_{2})\}$, 
independent of 
$\tau$ and  $\piop{1, \tau}{QIQ}$; details
are in the supplemental material.  
Denote $\piop{2, \tau}{QIQ}$ by
$\pitwoStar$.

Next we characterize
$\piop{1, \tau}{QIQ}$, which will motivate an
algorithm for calculating it. Let 
$d(\bh_{1}, y)$ be as in \eqref{dFn}, and define
$\Gam{\bh_1}{y} \triangleq \mbox{sgn}\{ d(\bh_{1}, y) \}$.  Then
$\Gam{\bh_1}{y} = \piop{1,
  \lambda}{TIQ}(\bh_1)|_{\lambda = y}$ 
is the optimal first-stage decision rule of TIQ-learning
at $\lambda = y$. We have introduced the new notation to 
emphasize the dependence on $y$. Next, define
the optimal
  $\tau^{\hbox{\scriptsize th}}$ quantile
\begin{equation}
\label{tstarDef}
\tStar \triangleq \inf \left \{y:
  \prOT{\Gam{\cdot}{y}}{\pitwoStar}(Y  
  \leq y)  \geq  \tau \right \},
\end{equation}
which we study further
  in the remainder of this section.

Lemma 7.7 of the supplemental material proves that 
$\lim_{y\rightarrow\infty (-\infty)}
\prOT{\Gam{\cdot}{y}}{\pitwoStar}(Y  
  \leq y) = 1 (0)$, so that $\tStar$ is defined for all $\tau \in (0, 
  1)$. 
For each $y \in \mathbb{R}$,  
\begin{eqnarray*}
\prOT{\pione}{\pitwoStar}(Y \leq y ) &=& E \left ( I
  \left[y, \Fdot, \g{\bH_1}{\pi_{1}(\bH_1)}{ \{ }{ \} } \right]
\right) \\ 
&\geq& E \left ( I
  \left[y, \Fdot, \g{\bH_1}{\Gam{\bH_1}{y}}{\{}{\}}
    \right] \right) \\
&=& \prOT{\Gam{\cdot}{y}}{\pitwoStar}(Y \leq y),
\end{eqnarray*}
where $I(\cdot, \cdot, \cdot)$ is defined in \eqref{It}.  The last
equality follows because $\Gam{\bH_1}{y}$ minimizes $E  
\left (I\left[y, \Fdot, \g{\bH_1}{a_1}{\{}{\}} \right] \right)$ with respect
to $a_1$. Hence,  $\left \{y: \prOT{\Gam{\cdot}{y}}{\pitwoStar}(Y \leq 
  y) \geq \tau \right \} \subseteq \left \{y:
  \prOT{\pione}{\pitwoStar}(Y 
  \leq  y ) \geq \tau \right \}$,
and taking the
infimum on both sides gives the upper bound 
\begin{equation}
\label{upBd}
\tStar \geq
\qOT{\pione}{\pitwoStar}(\tau) \mbox{ for all $\pione$}.
\end{equation} 
Thus, a first-stage decision rule
$\pione$  
is optimal if  it induces a $\tau^{\hbox{\scriptsize th}}$ quantile
equal to the upper bound $\tStar$ when treatments are subsequently 
assigned according to
$\pitwoStar$, i.e., if $\qOT{\pi_{1}}{\pitwoStar}(\tau) =
\tStar$. 

We now discuss conditions that guarantee existence of a
$\pi_{1}$ such that $\qOT{\pi_{1}}{\pitwoStar}(\tau) =
\tStar$ and derive its form. 
The quantile obtained under  regime
$\pi =  \{\Gam{\cdot}{y}, \pitwoStar\}$ is
\begin{equation}
\label{foft}
f(y) \triangleq \qOT{\Gam{\cdot}{y}}{\pitwoStar}(\tau) = \inf \left
  \{ \widetilde{y}: 
  \prOT{\Gam{\cdot}{y}}{\pitwoStar}(Y \leq \widetilde{y}) \geq \tau \right 
\}.
\end{equation}
Thus, because it is a quantile and the bound in \eqref{upBd} applies,
$\prOT{\Gam{\cdot}{y}}{\pitwoStar}\{Y \leq f(y)\} \geq \tau$, and
$f(y) = \qOT{\Gam{\cdot}{y}}{\pitwoStar}(\tau) \leq \tStar$ 
for all $y$. 
Our main results depend on the following lemma, which is proved in the
supplemental material. 

\begin{lem}\label{lemma1} 
\begin{eqnarray}
\label{iterUp}
&(\mbox{A})& y < \tStar \mbox{ implies } y < f(y) \leq
\tStar; \\ 
\label{leftlim}
&(\mbox{B})& f(\tStar^{-}) \triangleq \lim_{\delta
  \downarrow 0} f (\tStar - 
\delta) = \tStar; \\
&(\mbox{C})& f (\tStar) \leq \tStar \mbox{ with strict
  inequality if there exists } \delta > 0 \mbox{ such that } \nonumber
\\ 
\label{subDelt}
&& \prOT{\Gam{\cdot}{\tStar}}{\pitwoStar}(Y \leq 
  \tStar - \delta) \geq \tau; \\
\label{suffCondF}
&(\mbox{D})& \mbox{If $\Fdot$ is continuous and strictly increasing, 
  then $f(\tStar) = \tStar$}.
\end{eqnarray}
\end{lem}

It follows from $(B)$ that $f(\tStar) = \tStar$ if and only if $f(y)$ is left
continuous at  $y = \tStar$, and part $(D)$ is a sufficient  
condition guaranteeing left-continuity of $f(y)$ at $\tStar$. In this
case, the optimal first-stage 
rule is  $\piop{1,
  \tau}{QIQ}(\bh_{1}) = \Gam{\bh_{1}}{\tStar}$, i.e.,
$\qOT{\Gam{\cdot}{\tStar}}{\pitwoStar}(\tau) = 
\tStar$. The
condition stated in $(D)$ is commonly satisfied, e.g., when the 
density of
$\epsilon$ has positive support on the entire real line. If $f(y)$ is
not 
left continuous 
at $\tStar$, and thus $f(\tStar) < \tStar$, in light of
\eqref{leftlim} we can always approach the optimal policy via a
sequence of regimes of the form $\{ \Gam{\cdot}{\tStar - \delta_{n}},
\pitwoStar\}$, where $\delta_{n}$ decreases to 0. If the underlying 
distributions of the histories and $Y$ were known, the 
following algorithm   
produces an optimal regime.

\vspace{3mm}

\underline{\bf Population-level algorithm to find
  $\piop{1, \tau}{QIQ}$:}

\vspace{-2mm}

\begin{enumerate}[leftmargin=.65in]
\item[1.] Compute $\tStar$ from \eqref{tstarDef} and $f(\tStar)$ 
from \eqref{foft}. 

\item[2.] 
\begin{enumerate}
\item[a.] If $f(\tStar) = \tStar$, $\piop{1, \tau}{QIQ}(\bh_{1}) =
\Gam{\bh_1}{\tStar}$ is optimal  
 as it attains the quantile $\tStar$.

\item[b.]  If $f(\tStar) < \tStar$, $\lim_{\delta \downarrow 0}
\Gam{\bh_1}{\tStar - \delta}$ is optimal.
\end{enumerate}
\end{enumerate}


In practice, the generative model is not known, but the
population-level algorithm suggests an estimator of $\piop{1,
  \tau}{QIQ}$. The
following QIQ-learning algorithm can be used to 
estimate  an optimal
first-stage decision rule. The exact algorithm
depends on  
the choice of estimators for $\Fdot$ and $\g{\bh_1}{a_1}{(}{)}$;
several options are presented in Sections \ref{subsec:F} and
\ref{subsec:g}, but the choice should be data-driven; 
see, e.g., Remark \ref{parNonPar}.

\underline{\bf QIQ-learning algorithm:} 

\vspace{-2mm}

\begin{enumerate}[leftmargin=.82in]
\item[QIQ.1] Follow TIQ.1 -- TIQ.3 of the
  TIQ-learning algorithm in 
Section \ref{subsec:TIQLearning}.

\item[QIQ.2] With $I(\cdot, \cdot, \cdot)$ as in \eqref{It} and first-stage
  patient  histories $\bh_{1i}$, estimate
$\tStar$ using  
\begin{equation*}
\tStarHat \triangleq \inf \left ( y: \frac{1}{n} \sum_{i=1}^{n} I
  \left [ y, \FdotHat, \hatg{\bh_{1i}}{\hatGam{\bh_{1i}}{y}}{\{}{\}}
  \right ] \geq \tau
\right ).
\end{equation*}

\item[QIQ.3]  Estimate $f(\tStar)$ using
\begin{equation*}
\wh{f} (\tStarHat) \triangleq \inf \left ( y: \frac{1}{n}
  \sum_{i=1}^{n} I 
  \left [ y, \FdotHat,
    \hatg{\bh_{1i}}{\hatGam{\bh_{1i}}{\tStarHat}}{\{}{\}} 
   \right ] \geq \tau
\right ).
\end{equation*} 

\item[QIQ.4] 
\begin{enumerate}
\item[a.] If $\wh{f}(\tStarHat) = \tStarHat$, then
  $\hatpiop{1, \tau}{QIQ}(\bh_{1}) = \hatGam{\bh_{1}}{\tStarHat}$ is
  an  estimated optimal first-stage 
  decision rule  because it
  attains the estimated optimal quantile, $\tStarHat$.  
\item[b.] If $\wh{f}(\tStarHat) < \tStarHat$, then the first-stage
  rule $\pi_{1}(\bh_{1}) = 
 \hatGam{\bh_{1}}{\tStarHat-\delta}$, $\delta > 0$, 
results in the estimated quantile $\wh{f}(\tStarHat - \delta)$, which
satisfies $\tStarHat - \delta < \wh{f}(\tStarHat - \delta) \leq
\tStarHat$. By choosing $\delta$ arbitrarily small, this 
estimated quantile will be arbitrarily close to the estimated optimal  
quantile $\tStarHat$. 
\end{enumerate}
\end{enumerate}

To complete the TIQ- and
QIQ-learning  
algorithms,  we provide specific estimators $\Fdot$ and
$\g{\bh_1}{a_1}{(}{)}$ in the next two 
sections. We 
suggest estimators that are likely to be useful in practice, but our
list is not exhaustive. An advantage of TIQ-
and QIQ-learning is that they involve modeling 
only  smooth transformations of the data; these are standard, 
well-studied modeling problems in the statistics literature.  

\subsection{Working models for $\Fdot$}
\label{subsec:F}

Both TIQ- and QIQ-learning  require estimation of the 
distribution function of the second-stage error,
$\epsilon$.  We suggest two estimators that are
  useful in practice. The choice between them can be   
guided by inspection of the residuals from the second-stage
regression.  

\vspace{2mm}

\underline{\bf Normal Scale Model.}

The normal scale estimator for $\Fdot$ is 
$\wh{F}^{N}_{\epsilon}(z) \triangleq \Phi \left (
  z/\wh{\sigma}_{\epsilon} \right )$, where $\Phi(\cdot)$ denotes the
standard normal distribution function and $\wh{\sigma}_{\epsilon}$ 
is the standard deviation of the second-stage residuals, 
$\wh{e}_{i}^{Y} \triangleq Y_{i} - m(\bH_{2i}) -
A_{2i}c(\bH_{2i})$, $i=1, \dots, n$. 
If it is thought that $\sigma_{\epsilon}$ depends on
  $(\bH_{2}, A_{2})$, flexibility can be gained by
assuming a heteroskedastic variance model \citep{Car+Rup:88}, i.e.,
by assuming $F_{\epsilon}(z) = \Phi\left\lbrace 
z/\sigma_{\epsilon}(\bH_{2}, A_2)\right\rbrace$ for some unknown
function $\sigma_{\epsilon}(\bh_2, a_2)$.  
Given an estimator $\widehat{\sigma}_{\epsilon}(\bh_2, a_2)$ of 
$\sigma_{\epsilon}(\bh_2, a_2)$, an estimator of $\Fdot$ is
$\wh{F}^{N}_{\epsilon}(z) \triangleq \Phi \left \{
  z/\wh{\sigma}_{\epsilon}(\bH_{2}, A_{2}) \right \}$. We discuss
variance modeling techniques in the next section.

\vspace{2mm}
\underline{\bf Nonparametric Model.} 
For more flexibility, a non- or semi-parametric estimator
for $\Fdot$ can be used.  In the homogeneous variance case,
a nonparametric estimator of $\Fdot$ 
is the empirical distribution of the residuals, 
$\wh{F}^{E}_{\epsilon}(z)
\triangleq {n}^{-1} \sum_{i=1}^{n}
\mathbbm{1}(\wh{e}_{i}^{Y} \leq z)$.  In the heterogeneous variance
case, 
one can assume a non- or semi-parametric scale model 
$F_{\epsilon|\hbox{\scriptsize $\bH_2$}, A_2}(z|\bH_2=\bh_2, A_2=a_2) = F_{0}
\left\lbrace z/\sigma_{\epsilon}(\bh_2, a_2)\right\rbrace$, where
$F_{0}(\cdot)$ is an unspecified distribution function.  Given
an estimator $\widehat{\sigma}_{\epsilon}(\bh_2, a_2)$ 
of $\sigma_{\epsilon}(\bh_2, a_2)$, an estimator of 
$F_{\epsilon|\hbox{\scriptsize $\bH_2$}, A_2}(z|\bH_2=\bh_2, A_2=a_2)$ is
$\widehat{F}_{\epsilon}^{E}(z|\bH_2=\bh_2, A_2=a_2) = 
n^{-1}\sum_{i=1}^{n}\mathbbm{1}\left\lbrace
\widetilde{e}^{Y}_{i} \le z/\widehat{\sigma}_{\epsilon}(\bh_2, a_2)\right\rbrace$,
where $\widetilde{e}^{Y}_{i} = 
\widehat{e}_{i}^{Y}/\widehat{\sigma}_{\epsilon}(\bH_{2i}, A_{2i})$.  
Standard residual diagnostic techniques, e.g., a normal quantile-quantile plot,
 can be used to determine  
whether a normal assumption seems plausible for the observed
data.


\subsection{Working models for $\g{\bh_1}{a_1}{(}{)}$}
\label{subsec:g}

In addition to a model for $\Fdot$, TIQ- and
QIQ-learning require models for the bivariate
conditional density of 
$m(\bH_{2})$ and $c(\bH_{2})$ given $\bH_{1}$ and $A_{1}$.  A useful
strategy is to first
model the conditional mean and variance functions of $m(\bH_{2})$ and  
$c(\bH_{2})$ and then estimate the joint distribution of their
standardized residuals. Define these standardized residuals as
\begin{eqnarray*}
e^{m} = \frac{m(\bH_{2}) - \mu_{m}(\bH_{1},
  A_{1})}{\sigma_{m}(\bH_{1},A_{1})}, &&
e^{c} = \frac{c(\bH_{2}) - \mu_{c}(\bH_{1},
  A_{1})}{\sigma_{c}(\bH_{1},A_{1})},
\end{eqnarray*}
where $\mu_{m}(\bH_{1},A_{1}) \triangleq E\{m(\bH_{2}) \mid
\bH_{1},A_{1}\}$ and $\sigma_{m}^{2}(\bH_{1}, A_{1})
\triangleq E[\{m(\bH_{2}) - \mu_{m}(\bH_{1},A_{1})\}^{2} \mid
\bH_{1},A_{1}]$.
The mean and variance functions of $c(\bH_{2})$ are
defined similarly: $\mu_{c}(\bH_{1}, A_{1}) \triangleq
E\{c(\bH_{2}) | \bH_{1},\allowbreak A_{1}\}$, and 
$\sigma_{c}^{2} (\bH_{1}, A_{1}) \triangleq E[\{c(\bH_{2}) -
\mu_{c}(\bH_{1},A_{1})\}^{2} | \bH_{1},A_{1}]$.  In simulations, we use
parametric mean and variance models for $\mu_{m}$, $\sigma_{m}^{2}$,   
$\mu_{c}$, and $\sigma_{c}^{2}$, and we estimate the joint
distribution of 
$e^m$ and $e^c$ using a Gaussian copula. Alternatively, the joint
residual distribution could be modelled parametrically, e.g., 
with a multivariate normal model; or nonparametrically, e.g., using a
bivariate kernel density estimator  
\citep[][Ch. 4]{Sil:86}. The Gaussian
copula is used in the simulations in Section
\ref{sec:montecarloresults}, and results are 
provided using a bivariate kernel estimator in the   
supplemental material. Common exploratory analysis
techniques can be 
used to interactively guide the choice of estimator for
$\g{\bh_{1}}{a_{1}}{(}{)}$. 
In simulated experiments described in the supplemental material,
a bivariate kernel density estimator was competitive with
a correctly specified Gaussian copula model with sample sizes
as small as $n=100$. Using parametric mean and variance modeling, the
following steps  would
be substituted in Step TIQ.3 of the TIQ-learning
algorithm. 

\vspace{3mm}

\underline{\bf Mean and Variance Modeling.}

\begin{enumerate}[leftmargin=.6in]
\item[3.1] Compute $\wh{\btheta}_{m} \triangleq
  \arg \min_{\hbox{\scriptsize $\btheta_{m}$}}  
  \sum_{i=1}^{n} \left \{\widehat{m}(\bH_{2i}) -
    \mu_{m}(\bH_{1i}, A_{1i}; \btheta_{m})\right \}^{2}$ and the
  resulting estimator $\mu_{m}(\bH_{1}, A_{1}; \wh{\btheta}_{m})$ of
  the mean function $\mu_{m}(\bH_{1}, A_{1})$.
  
\item[3.2]  Use the estimated mean function from Step 3.1 to obtain    
\begin{equation*}
\wh{\bgamma}_{m} \triangleq \argmin{\hbox{\scriptsize $\bgamma_{m}$}}
\sum_{i=1}^{n} 
 \left[ \left \{\widehat{m}(\bH_{2i}) - 
    \mu_{m}(\bH_{1i},A_{1i}; \wh{\btheta}_{m})\right \}^{2} -
  \sigma_{m}^{2}(\bH_{1i}, A_{1i}; \bgamma_{m}) \right]^{2},
\end{equation*}
and subsequently the estimator
 $\sigma_{m}^{2}(\bH_{1}, A_{1}; \wh{\bgamma}_{m})$
of $\sigma_{m}^{2}(\bH_{1}, A_{1})$.
One choice
for $\sigma_{m}(\bh_1, a_2; \gamma_{m})$ is a log-linear model,
which may include non-linear basis terms.  

\item[3.3] Repeat Steps 3.1 and 3.2 to obtain estimators 
$\mu_{c}(\bH_1, A_1; \widehat{\btheta}_{c})$ and 
$\sigma_{c}(\bH_1, A_1;
\widehat{\bgamma}_{c})$. 

\item[3.4]  
  Compute standardized residuals
  $\wh{e}_{i}^{m}$ and $\wh{e}_{i}^{c}$, $i = 1, ..., n$,
  as
\begin{eqnarray*}
\wh{e}^{m}_{i} = \frac{\widehat{m}(\bH_{2i}) -
  \mu_{m}(\bH_{1i}, A_{1i};
  \wh{\btheta}_{m})}{\sigma_{m}(\bH_{1i},A_{1i};
  \wh{\bgamma}_{m})}, && 
\wh{e}^{c}_{i} = \frac{\widehat{c}(\bH_{2i}) -
  \mu_{c}(\bH_{1i}, A_{1i};
  \wh{\btheta}_{c})}{\sigma_{c}(\bH_{1i},A_{1i};
  \wh{\bgamma}_{c})}.
\end{eqnarray*}
\end{enumerate}
Then, $\wh{e}^{m}_{i}$ and $\wh{e}^{c}_{i}$, $i = 1, ..., n$, can be
used to 
estimate the joint distribution of the standardized residuals.
Samples drawn from this distribution can be transformed back to
samples from $\hatg{\bh_{1}}{a_{1}}{(}{)}$ to estimate the 
integral $I\left\{ y, \FdotHat, \hatg{\bh_1}{a_1}{(}{)} \right\}$
with a Monte Carlo average. 

\subsection{Theoretical results}
\label{subsec:theory}

The following assumptions are used to
establish consistency of the threshold exceedance
probability and quantile that result from applying the estimated 
TIQ- and QIQ-learning optimal regimes, 
respectively.  For each $\bh_1, a_1,$ and $\bh_2$:

\vspace{2mm}


\noindent A1. the method used to estimate $m(\cdot)$ 
and $c(\cdot)$ results in estimators 
$\wh{m}(\bh_{2})$ and 
$\wh{c}(\bh_{2})$ that converge in probability to
$m(\bh_{2})$ and 
$c(\bh_{2})$, respectively;   

\noindent A2. $\Fdot$ is continuous, $\FdotHat$ is a cumulative 
distribution function, and 
$\FepsHat(y)$ converges in probability to $\Feps(y)$ uniformly in
$y$;

\noindent A3.  $\int |d\wh{G}(u,
v \mid  
\bh_{1}, a_{1}) - dG (u, v \mid \bh_1, a_1)|$ converges to zero in
probability;

\noindent A4.  $n^{-1}\sum_{i=1}^{n}\int |
d\wh{G}(u, v \mid \bH_{1i}, a_{1}) - dG (u, v \mid \bH_{1i}, a_{1})|$ 
converges to zero in probability;

\noindent A5. $\mbox{pr}\{|d( \bH_{1}, \tStar)| = 0 \} = 0$. \\

\vspace{-4mm}

\noindent In the simulation experiments in Section
\ref{sec:montecarloresults} and data example in Section
\ref{sec:stard}, we use linear working models for $m(\cdot)$ and 
$c(\cdot)$ that are estimated using least squares. Thus, A1 is
satisfied under usual regularity conditions. When $\epsilon$ is
continuous, 
assumption A2 can be satisfied by specifying $\FdotHat$ as the
empirical 
distribution function. If for each fixed $\bh_1$ and $a_1$, $dG (\cdot,
\cdot 
\mid \bh_1, a_1)$ is a 
density and $d\wh{G}(\cdot, \cdot \mid
\bh_{1}, a_{1})$ a pointwise consistent estimator, then A3 is
satisfied 
\citep{Gli:74}. 
Assumption A5 states that all patients have a non-zero first-stage 
treatment 
effect at $\tStar$. Theorems \ref{tiqThm} and \ref{qiqThm} are proved in the
supplemental material.  

\begin{thm}
\label{tiqThm}
(Consistency of TIQ-learning) Assume A1--A3 and  
fix $\lambda \in \mathbb{R}$. Then,
$\bprHatpiop{\lambda}{TIQ}(Y>\lambda)$  converges in
probability to $\bprpiop{\lambda}{TIQ}(Y>\lambda)$, where
$\bhatpiop{\lambda}{TIQ} = (\hatpiop{1, \lambda}{TIQ},
\hatpitwoStar)$. 
\end{thm}

\begin{thm}
\label{qiqThm}
(Consistency of QIQ-learning) Assume
A1--A5. Then, 
$\bqHatpiop{\tau}{QIQ}(\tau)$ converges in probability to
$\tStar$ for any fixed $\tau$, where
$\bhatpiop{\tau}{QIQ} =
(\hatGam{\cdot}{\tStarHat},  \hatpitwoStar)$.
\end{thm}

\section{Simulation experiments}
\label{sec:montecarloresults}

We compare the performance of our estimators to binary $Q$-learning
\citep{Cha+Moo:13}, $Q$-learning, and  
the mean-optimal method IQ-learning
\citep{Lab+etal:13} for a range of data generative models.  Gains are
achieved in terms of the proportion of the distribution of $Y$ that
exceeds the constant threshold $\lambda$ and the
$\tau^{\hbox{\scriptsize th}}$ quantile for several values of
$\lambda$ and $\tau$.  The data are generated using the model 
\begin{equation*}
\begin{array}{lll}
\bX_{1} \sim \mbox{Norm}({\bf 1}_{2}, {\boldsymbol \Sigma}), 
& A_{1}, A_2 \sim \mbox{Unif}\{ -1, 1\}^2, &  \bH_{1} = (1,
\bX_{1}^{\T})^{\T}, \\ 
\eta_{\hbox{\scriptsize $\bH_{1}$}, A_{1}} = \exp \{ \frac{C}{2}(\bH_{1}^{\T}\bgamma_{0} +
A_{1}\bH_{1}^{\T}\bgamma_{1}) \}, & 
{\boldsymbol \xi} \sim
\mbox{Norm}( {\bf 0}_{2}, {\bf I}_{2}), &
\bX_{2} = \bB_{A_{1}}\bX_{1} + \eta_{\hbox{\scriptsize $\bH_{1}$},
  A_{1}}{\boldsymbol \xi}, \\ \bH_{2} = (1, \bX_{2}^{\T})^{\T}, &
\epsilon \sim \mbox{Norm}(0,1), & 
Y = \bH_{2}^{\T}\bbeta_{2,0} + A_{2}\bH_{2}^{\T}\bbeta_{2,1} + \epsilon,
\end{array}
\end{equation*}
where ${\bf 1}_p$ is a $p \times 1$ vector of 1s, $\mathbf{I}_{q}$ is
the $q\times q$ identify matrix, and $C\in[0,1]$ is
a constant. The matrix
${\boldsymbol \Sigma}$ is a  
correlation matrix with off-diagonal
$\rho=0.5$. The $2 \times 2$ matrix $\bB_{A_{1}}$ equals 
\begin{eqnarray*}
\bB_{A_{1}=1} = \begin{pmatrix}
 -0.1 & -0.1 \\
0.1 & 0.1
\end{pmatrix}, && \bB_{A_{1}=-1} = \begin{pmatrix}
0.5, & -0.1 \\
-0.1 & 0.5
\end{pmatrix}.
\end{eqnarray*}
The remaining parameters are $\bgamma_{0} = (1, 0.5, 0)^{\T},$
$\bgamma_{1} = (-1, -0.5, 0)^{\T}$, $\bbeta_{2,0} = (0.25, -1,
0.5)^{\T}$,  and $\bbeta_{2,1} = (1, -0.5, -0.25)^{\T}$, 
which were chosen to ensure 
that the mean-optimal
treatment produced a more variable response for some patients.  
 \begin{figure}
 \begin{center}
   \includegraphics[scale=.55]{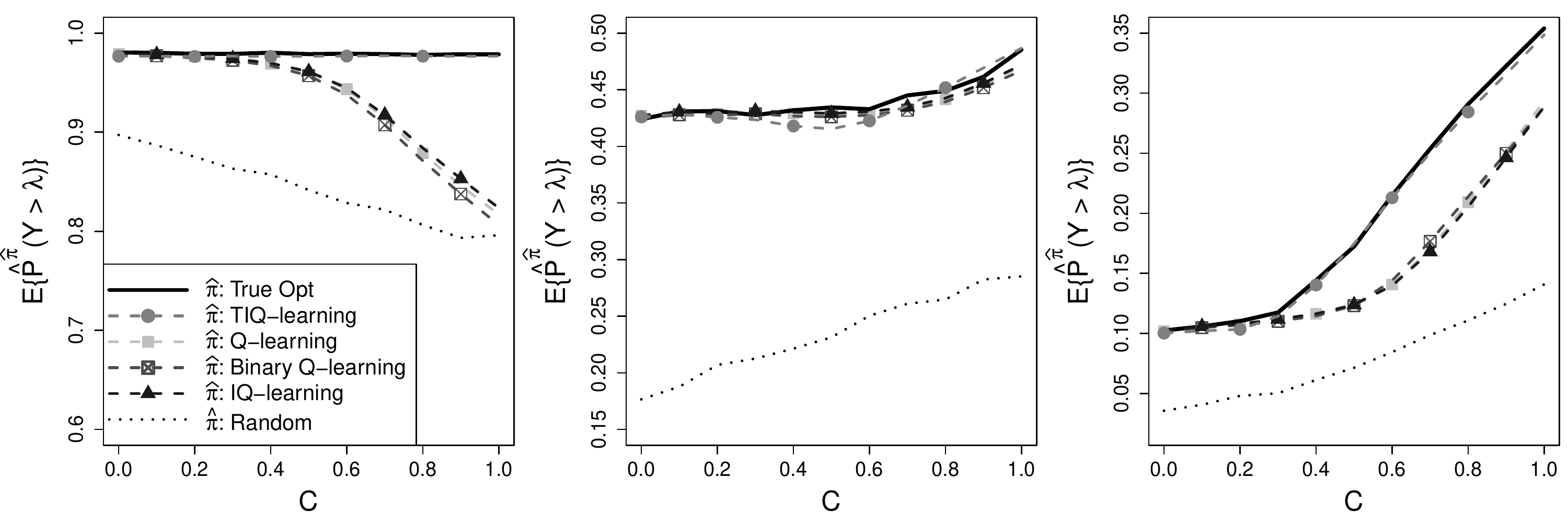}
   \caption{\emph{Left to Right:} $\lambda = -2, 2, 4$. Solid black,
     true  
     optimal threshold probabilities; dotted black, probabilites under
     randomization; dashed with circles/squares/crossed squares/triangles,
     probabilities under TIQ-, $Q$-, binary $Q$-, and Interactive 
     $Q$-learning, respectively.}\label{tiqSim1}
 \end{center}
 \end{figure} 
\subsection{TIQ-learning Simulation Results} 
Results are based on $J=1,000$ generated data sets; for each, 
we estimate the TIQ-, IQ-, binary $Q$-learning, and 
$Q$-learning policies using a training set of size $n=250$
and compare the results using a test set of size $N=10,000$.   The
normal 
scale model is used to estimate $\Fdot$, which is correctly
specified for the generative model above. The Gaussian copula model 
discussed in Section 2.4 is also correctly specified and is used as
the 
estimator for $\g{\bh_{1}}{a_{1}}{(}{)}$.  Results using a bivariate
kernel estimator for $\g{\bh_{1}}{a_{1}}{(}{)}$ are presented in the
  supplemental material.

To study the performance of the TIQ-learning 
algorithm, we compare 
values of the cumulative distribution function of the final response
when treatment is assigned according to the estimated
TIQ-learning, IQ-learning, binary $Q$-learning, and
$Q$-learning regimes.  Define $\mbox{pr}^{\hbox{\scriptsize
    $\wh{\bpi}_{j}$}}(Y > 
\lambda)$ to be 
the true probability that $Y$ exceeds 
$\lambda$ given treatments are assigned according to
$\wh{\bpi}_{j} = (\wh{\pi}_{1j}, \wh{\pi}_{2j})$, the regime estimated
from the 
$j^{\hbox{\scriptsize th}}$  
generated data set. For threshold values
$\lambda = -2, 2, 
4$, we estimate $\mbox{pr}^{\hbox{\scriptsize $\bpi$}}(Y > \lambda)$
using 
$\sum_{j=1}^{J} 
\wh{\mbox{pr}}^{\hbox{\scriptsize $\wh{\bpi}_{j}$}}(Y > \lambda)/J$,
where   
$\wh{\mbox{pr}}^{\hbox{\scriptsize $\wh{\bpi}_{j}$}}(Y > \lambda)$ is
an estimate of 
$\mbox{pr}^{\hbox{\scriptsize $\wh{\bpi}_{j}$}}(Y 
> \lambda)$ obtained by calculating the proportion
of test patients consistent with regime $\wh{\bpi}_{j}$ whose observed  
$Y$ 
values are greater than $\lambda$. Thus, our estimate is an
average over training data sets and test set observations. In terms
of the proportion of distribution mass above $\lambda$, results for
$\lambda=-2$ and 4 in
Figure \ref{tiqSim1} show a clear 
advantage of TIQ-learning for higher values of $C$, the degree 
of heteroskedasticity in the second-stage covariates $\bX_{2}$.  As
anticipated by Remark \ref{diffOpt} in Section 2.1, all methods
perform similarly when $\lambda=2$.  
 \begin{figure}
 \begin{center}
   \includegraphics[scale=.53]{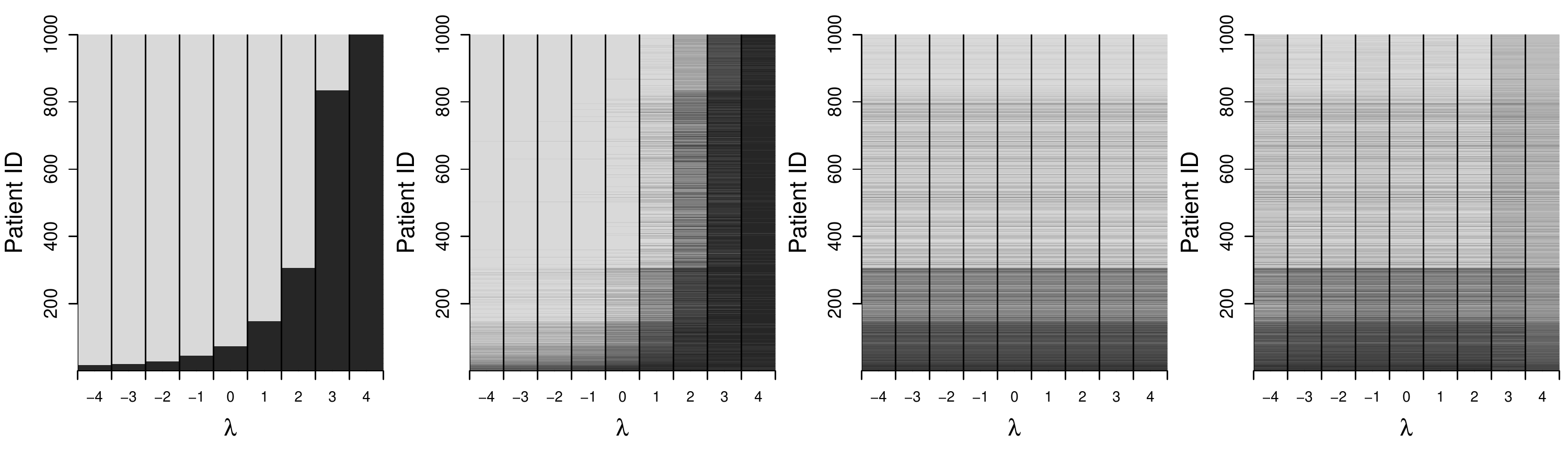}
   \caption{\emph{From left:} True optimal first-stage treatments for 1,000 
     test 
     set patients when $\lambda = -4, -3,..., 4$, coded light
     gray when $\piop{1, \lambda}{TIQ}(h_1)=1$ and dark gray
       otherwise;
   TIQ-learning estimated optimal first-stage
   treatments; 
   $Q$-learning estimated optimal first-stage treatments, plotted 
   constant in 
   $\lambda$ to aid visual comparison; and binary $Q$-learning
   estimated optimal first-stage treatments for each $\lambda$.}\label{tiqPol1} 
 \end{center}
 \end{figure}

Figure \ref{tiqPol1} illustrates how the optimal first-stage treatment for 
a test set of 1,000 individuals changes as $\lambda$ varies. Results 
are shown for $C=0.5$. The true optimal
treatments displayed in the left plot show a distinct shift from
treating most of the 
population with $A_{1}=1$ to $A_{1}=-1$ as $\lambda$ 
increases from -4 to 4. The TIQ-learning
estimated optimal 
treatments displayed in the middle 
plot are averaged over 100 Monte Carlo iterations and closely resemble 
the true policies on the left. 
Although the  estimated
$Q$-learning regime does not depend on $\lambda$, it is 
plotted for each $\lambda$ value to aid visual comparison. The
first-stage treatments recommended by $Q$-learning differ the most
from the true 
optimal treatments when $\lambda = 4$, corroborating the results for
$C = 0.5$ in Figure \ref{tiqSim1}. The rightmost panel of Figure
\ref{tiqPol1} are the results from binary $Q$-learning with the binary
outcome defined as $\mathbbm{1}_{Y > \lambda}$. While there appears to
be a slight deviation in the results from mean-optimal
$Q$-learning for $\lambda$ values 3 and 4, overall the resulting
policies are similar to mean-optimal $Q$-learning and do not recover
the true optimal treatments on average. 

\subsection{QIQ-learning Simulations}
\label{subsec:qiqresults}

To study the performance of the QIQ-learning algorithm, we compare 
quantiles of $Y$ when the population is
treated according to the regimes estimated by QIQ-learning,
IQ-learning, and $Q$-learning.  A smaller test set of
size $N = 5,000$ was used in this section to reduce computation
time. Define 
$q^{\hbox{\scriptsize $\wh{\bpi}_{j}$}}(\tau)$ to be the true
$\tau^{\hbox{\scriptsize th}}$ quantile of the distribution of $Y$
given treatments are assigned according to 
$\wh{\bpi}_{j} = (\wh{\pi}_{1j}, \wh{\pi}_{2j})$, the regime estimated
from the $j^{\hbox{\scriptsize th}}$ 
generated data set. For $\tau = 
0.1, 0.5, 0.75$, we estimate $q^{\hbox{\scriptsize
    $\bpi$}}(\tau)$ using 
$\sum_{j=1}^{J} 
\wh{q}^{\hbox{\scriptsize $\wh{\bpi}_{j}$}}(\tau)/J$, where  
$\wh{q}^{\hbox{\scriptsize $\wh{\bpi}_{j}$}}(\tau)$ is an estimate of
$q^{\hbox{\scriptsize $\wh{\bpi}_{j}$}}(\tau)$ obtained by  calculating the
$\tau^{\hbox{\scriptsize th}}$ quantile of the subgroup 
of test patients consistent with regime $\wh{\bpi}_{j}$. The
generative model  and all other parameter settings used
here are the same as those in the previous section. For our generative 
model, the condition of Lemma \ref{lemma1} is satisfied, so the true
optimal regime is attained asymptotically.  
 \begin{figure}
 \begin{center}
 \includegraphics[scale=.6]{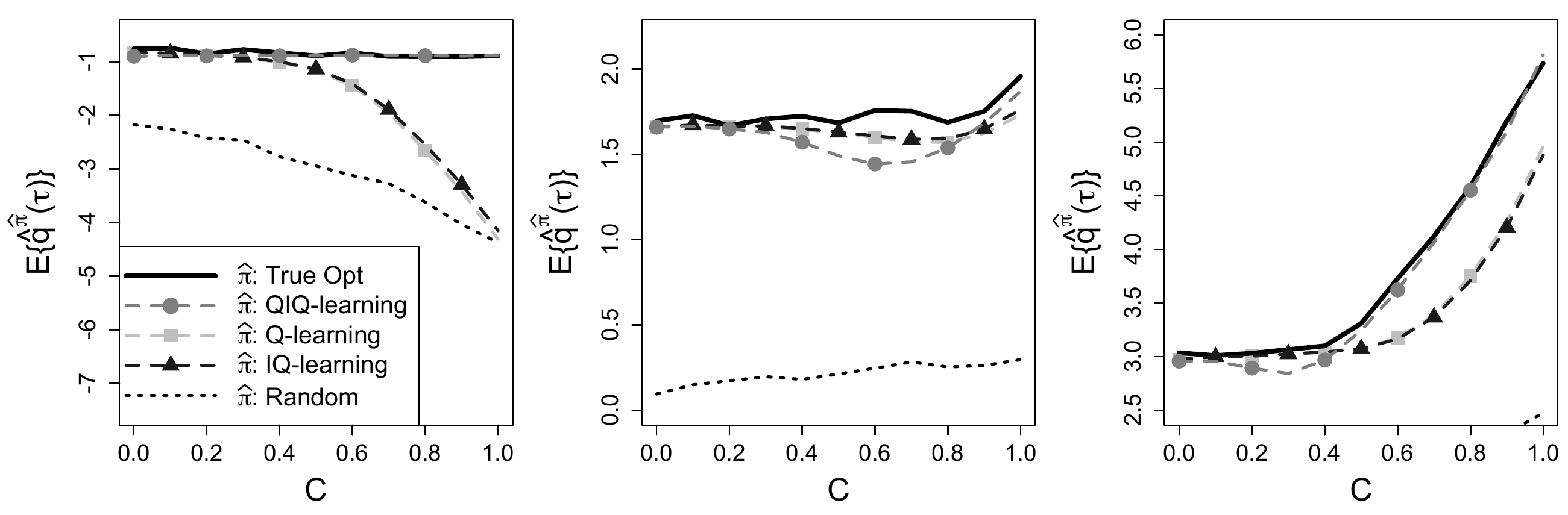}
 \caption{\emph{Left to Right:} $\tau = 0.1, 0.5, 0.75$. Solid black, 
     true optimal quantiles; dotten black, quantiles under
     randomization; dashed with circles/squares/triangles, 
     quantiles under QIQ-, $Q$-, and
     IQ-learning, respectively.}\label{qiqSim1}
 \end{center}
 \end{figure}
The results in Figure \ref{qiqSim1} indicate that the lowest quantile,
$\tau = 
0.1$, suffers under the $Q$-learning regime as heterogeneity in
the second-stage histories increases, measured by the scaling constant
$C$.  In  contrast, 
quantiles of the QIQ-learning estimated regimes
for $\tau=0.1$ remain 
constant across the entire range of $C$. When $\tau=0.5$, all methods 
perform similarly; 
for some $C$, IQ- and $Q$-learning outperform
QIQ-learning. This is not surprising because all 
models used 
to generate the data were 
symmetric. Thus, maximizing the mean of $Y$ gives similar results to
maximizing the median.  


Next we study QIQ-learning when the first stage errors are skewed. 
The generative model and parameter settings used
here are the same as those used previously except that 
\begin{equation*}
\begin{array}{ll}
\bX_{2} = \bB_{A_{1}}\bX_{1} + \eta_{\hbox{\scriptsize $\bH_{1}$},
  A_{1}}{\boldsymbol \xi}, & 
\eta_{\hbox{\scriptsize $\bH_{1}$}, A_{1}} = \exp \{ \frac{1}{2}(\bH_{1}^{\T}\bgamma_{0} +
A_{1}\bH_{1}^{\T}\bgamma_{1}) \}, \\
{(10C + 1) + \sqrt{2(10C+1)}\boldsymbol \xi} \sim
\chi^{2}_{df = 10C+1}, 
\end{array}
\end{equation*}
where  $C\in[0,1]$ is
a constant that reflects the degree of skewness in the first-stage
errors, ${\boldsymbol \xi}$. Smaller values of $C$ correspond to
heavier skew. 
 \begin{figure}
 \begin{center}
   \includegraphics[scale=.53]{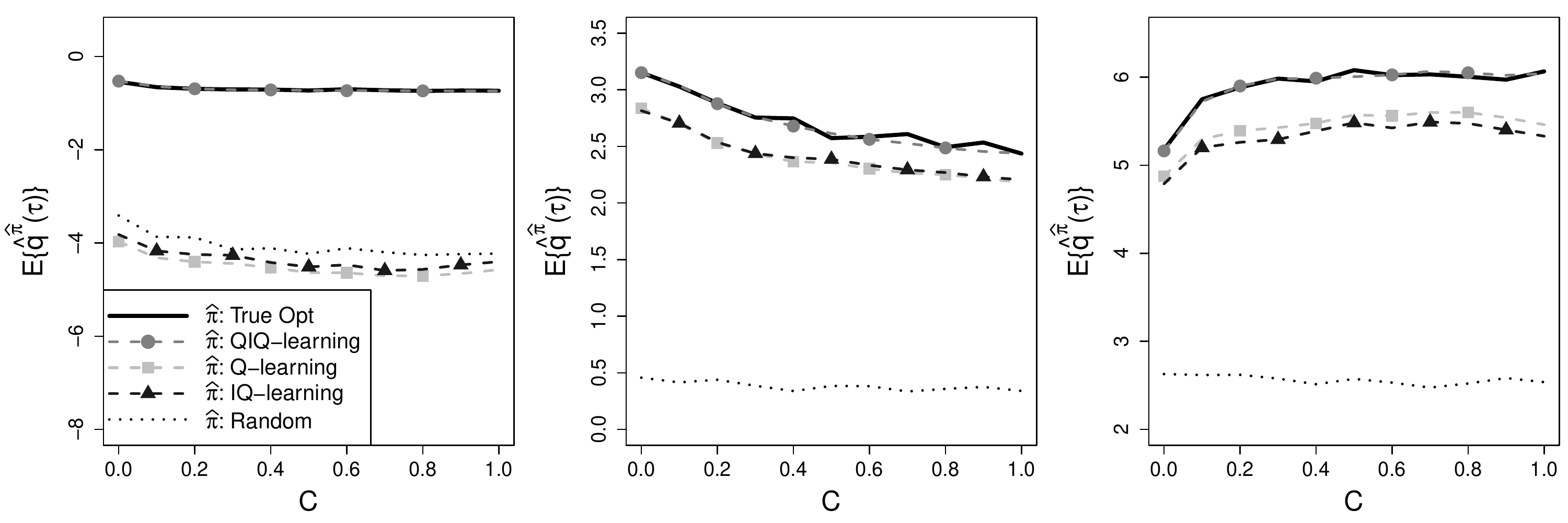}
   \caption{\emph{Left to Right:} $\tau = 0.1, 0.5, 0.75$. Solid black,
     true  
     optimal threshold probabilities; dotted black, probabilites under
     randomization; dashed with circles/squares/triangles,
     probabilities under TIQ-, $Q$-, and Interactive 
     $Q$-learning, respectively. Training set size of $n=500$.}\label{skew}
 \end{center}
 \end{figure}  

Results are averaged over $J=100$ generated data sets; for each, 
we estimate the QIQ-, IQ-, and 
$Q$-learning policies
and compare the results using a test set of size $N=10,000$.   
The training sample size for each iteration is $n=500$. 
The normal 
scale model is used to estimate $\Fdot$, which is correctly
specified. A bivariate kernel density
estimator is used to estimate $\g{\bh_{1}}{a_{1}}{(}{)}$.  
As before, we compare 
quantiles of the final response
when treatment is assigned according to the estimated
QIQ-learning, IQ-learning, and
$Q$-learning regimes, and the results are given in Figure \ref{skew}.  
QIQ-learning demonstrates an advantage over the
mean-optimal methods for all three quantiles and almost uniformly
across the degree of skewness of the first-stage errors.

\section{STAR*D Analysis}
\label{sec:stard}

The Sequenced Treatment Alternatives to Relieve Depression (STAR*D)
trial \citep{Fav+etal:03, Rus+etal:04} is a four-stage Sequential
Multiple Assignment Randomized Trial \citep{Lav+Daw:04, Mur:05b}
studying  personalized treatment strategies for patients with major
depressive disorder. Depression is measured by the 
Quick Inventory of Depressive Symptomatology (QIDS) score, a
one-number summary score that takes
integer values 0 to 27. Lower scores
indicate fewer depression symptoms. Remission is defined
as QIDS $\leq 5$.   Previous attempts to estimate
optimal dynamic treatment regimes from this data have used the
criteria, ``maximize end-of-stage-two QIDS," \citep[see, for
example,][]{Sch+etal:12, Lab+etal:13} a surrogate for the primary
aim of helping patients achieve remission. We illustrate TIQ-learning
by estimating an optimal regime that 
maximizes 
the probability of  
remission for each patient, directly corresponding to
the primary clinical goal.  

The first stage, which we
will henceforth refer to as baseline, was non-randomized with each
patient receiving Citalopram, a drug in the class of Selective
Serotonin Reuptake  
Inhibitors (SSRIs). We use a subset of the STAR*D data from the first
two randomized stages, and refer to the original trial levels 2 and 3
as ``stage one" and ``stage two." Before each
randomization, patients specified a preference to 
``switch''  or ``augment''  their current treatment strategy and were
than randomized to one of multiple options within their preferred
category.  In addition, patients who achieved remission
in any stage exited the study.  To keep our illustration
of TIQ-learning concise, we restrict our analysis to the subset of
patients who who preferred the ``switch''
strategy at both stages. We note that this subgroup is not
identifiable at baseline because patient preferences depend on the
assigned treatment and subsequent response at each stage. Our
motivation for this restriction is to mimic a two-stage SMART 
where treatments are randomized at both stages, thus simplifying our
illustration.
At stage one, our binary treatment variable 
is ``SSRI," which includes only Sertraline, versus ``non-SSRI," which
includes 
both Bupropion and Venlafaxine. At stage two we 
compare Mirtazapine and
Nortriptyline which are both non-SSRIs. In the patient subgroup 
considered in our 
analysis, treatments were randomized at both stages.

All measured QIDS scores are recoded as 
$27 - $ QIDS so that higher scores correspond to fewer 
depression symptoms.  After recoding, remission corresponds to QIDS 
$ > 21$. Thus, TIQ-learning with $\lambda = 21$  
maximizes 
the probability of remission for all patients.   In general, QIDS was
recorded during clinic 
visits at weeks 2, 4, 6, 9, and 12 in each stage, although some
patients with inadequate response moved on to the next stage before 
completing all visits. We summarize longitudinal QIDS trajectories
from 
the baseline stage and stage one by averaging over the total number of  
QIDS observations in the given stage. 
Variables used in our analysis are listed in Table
\ref{stardVariables2}.     
\begin{table}[here]
\caption{Variables used in the STAR*D
  analysis.}\label{stardVariables2} 
\vspace{2mm}
\begin{tabular}{c p{12.5cm}}
Variable & Description \\
\hline
  \texttt{qids0} & mean QIDS during the baseline
  stage. \\  
  \texttt{slope0} & pre-randomization QIDS improvement; the 
  difference between the final and initial baseline-stage QIDS scores,
  divided by time spent in the baseline stage. \\   
  \texttt{qids1}  & mean stage-one QIDS. \\  
  \texttt{slope1} & first-stage QIDS improvement; the 
  difference between the final and initial first-stage QIDS scores,
  divided by time spent in the first randomized stage. \\  
  \texttt{A1} & First-stage
  treatment; 1=``SSRI" and -1=``non-SSRI." \\ 
  \texttt{A2} & Second-stage
  treatment; 1=``NTP" for Nortriptyline and -1=``MIRT" for
  Mirtazapine. \\ 
  \texttt{Y} & 27 minus final QIDS score,
  measured at the end of stage two. \\ 
\end{tabular}
\end{table}
We describe all
models used in the analysis below.

At the second stage, we assume the linear working model $Y =
\bH_{2,0}^{\T}\bbeta_{2,0} +    
A_{2}\bH_{2,1}^{\T}\bbeta_{2,1} + \epsilon$,
where $\bH_{2,0} = \bH_{2,1} = (1, \texttt{ qids1},
\texttt{ slope1}, \texttt{ A1})^{\T}$, $E(\epsilon) 
=0$,  
var$(\epsilon) = \sigma^2$, and $\epsilon$ is 
independent of $\bH_{2}$ and $A_{2}$.  We fit this model using least 
squares. A normal qq-plot of the
residuals from the previous regression step indicates slight
deviation from normality, so we use the non-parametric estimator of
$\Feps(\cdot)$ given in Section 2.4. Next, we estimate the conditional
mean and variance functions of 
$m(\bH_{2}) \triangleq \bH_{2,0}^{\T}\bbeta_{2,0}$ and $c(\bH_{2})
\triangleq \bH_{2,1}^{\T}\bbeta_{2,1}$ following steps described in
Section 2.4. For the mean functions, we take $\bH_{1,0} = \bH_{1,1}=
(1, \bX_{1}^{\T})^{\T}$ 
with $\bX_{1} = (\texttt{qids0}, \texttt{ slope0})^{\T}$ and use
working models of the form 
$E\{ k(\bH_2) \mid \bX_{1}, A_{1}\} = \bH_{1,0}^{\T}\bbeta_{1,0}^{k} + 
A_{1}\bH_{1,1}^{\T}\bbeta_{1,1}^{k}$.
Exploratory analyses reveal little evidence of heteroskedasticity at the
first-stage. Thus, we opt to estimate a constant residual variance
for both terms following the mean modeling steps. After the mean
and variance modeling steps, we use a Gaussian copula 
to estimate the joint conditional distribution of the standardized
residuals of $\{m(\bH_{2}), c(\bH_{2})\}$ given $\bH_{1}$ and $A_{1}$, 
resulting in our estimate of $\g{\bh_1}{a_1}{(}{)}$ which we denote by
$\hatg{\bh_1}{a_1}{(}{)}$. 

The estimated first-stage optimal rule is $\hatpiop{1,
  \lambda}{TIQ}(\bh_{1}) = \argmin{a_{1}} \int \FepsHat  
  ( 21 - u - |v|) d\wh{G}(u,v \mid \bh_{1}, a_{1})$.
At stage two, $\hatpitwoStar(\bh_{2}) = \mbox{sgn}(-1.66 +
0.15*\mbox{\texttt{qids1}} -   
4.03*\mbox{\texttt{slope1}} - 0.68*\mbox{\texttt{A1}})$
is the estimated optimal treatment. Based on Remark 1 in Section
\ref{subsec:TIQLearning}, we 
compare the estimated  
first-stage treatment recommendations to those recommended by the     
mean-optimal rule, $\arg\min_{a_{1}} \int ( - u - |v|) d\wh{G}(u,v 
\mid \bh_{1}, a_{1})$, for each observed $\bh_{1}$ in the data. Only one
patient out of 132 is   
recommended differently.  In addition, the difference in raw values
of $\int \FepsHat ( 21 - u - |v|) d\wh{G}(u,v
\mid \bh_{1}, a_{1})$ for $a_{1} = 1, -1$ as well as $\int ( - u -
|v|) d\wh{G}(u,v  
\mid \bh_{1}, a_{1})$ for $a_{1} = 1, -1$ are the smallest for this
particular 
patient. Thus, the treatment discrepancy is most likely due to a
near-zero treatment effect for this patient.  
\begin{table}[here]
\vspace{10mm}
\caption{Estimated value of dynamic and non-dynamic
  regimes using the Adaptive Inverse Probability Weighted Estimator.}\label{value}
\begin{center}
\begin{tabular}{lccc}
& Estimated Value \\
\hline
TIQ-learning & 0.24 \\
$Q$-learning & 0.23 \\
Binary $Q$-learning & 0.19 \\
 (1, 1) & 0.13 \\
(-1, 1) & 0.24 \\
(1, -1) & 0.07 \\
(-1, -1) & 0.12 \\ 
\end{tabular}
\end{center}
\end{table}

We compare TIQ-learning to the $Q$-learning analysis of
  \citet{Sch+etal:12} and binary $Q$-learning \citep{Cha+Moo:13}.
Comparing the results to $Q$-learning, which
maximizes the expected 
value of $Y$, supports the claim that TIQ-learning and mean
optimization are equivalent for this subset of the STAR*D data. The
first step of $Q$-learning is to model the 
conditional expectation of $Y$ given $\bH_{2}$ and $A_{2}$ which is the
same as the first step of  TIQ-learning. Thus, we use the same model
and estimated decision rule at stage two given in Step 1 of the
TIQ-learning algorithm.  Next, we model the conditional expectation 
of $\widetilde{Y} = \bH_{2,0}^{\T}\bbeta_{2,0} +
|\bH_{2,1}^{\T}\bbeta_{2,1}|$, where $\widetilde{Y}$ is the predicted future
optimal outcome at stage one.  We specify the working model
$E(\widetilde{Y} \mid \bH_{1}, A_{1}) = \bH_{1,0}^{\T}\bbeta_{1,0}^{Q} +
A_{1}\bH_{1,1}^{\T}\bbeta_{1,1}^{Q}$,
where $\bH_{1,0}^{\T} = \bH_{1,1}^{\T} = (1, \bX_{1}^{\T})^{\T}$ and $\bX_{1}
= (\texttt{qids0}, \texttt{ slope0})^{\T}$. We fit the model using least
squares.  Then, the $Q$-learning
estimated optimal first-stage rule is  $\hatpiop{1,
  \lambda}{Q}(\bh_{1}) = \mbox{sgn}(-0.95 + 0.13*\mbox{\texttt{qids1}}
+  
2.17*\mbox{\texttt{slope1}})$.  $Q$-learning recommends treatment
differently at the first stage for only one of the 132 patients in the
data.  In addition, the estimated value of the TIQ- and
$Q$-learning regimes are nearly the same and are displayed in Table 
\ref{value}.  Binary $Q$-learning recommends treatment differently
than TIQ-learning for 18 patients at the first stage, and the estimated
value of the binary $Q$-learning regime is slightly lower than TIQ-
and $Q$-learning. Also included in Table \ref{value} are value estimates
for four   
non-dynamic  regimes  
that treat everyone according to the decision rules $\pi_{1}(\bh_{1})
= a_{1}$ and  $\pi_{2}(\bh_{2}) = a_{2}$ for $a_{1} \in \{-1, 1\}$ and
$a_{2} \in \{-1, 1\}$. We estimate these values using the Augmented
Inverse Probability Weighted Estimator given in \citet{Zha+etal:13}.  

In summary, it appears that TIQ-learning and $Q$-learning perform
similarly for this 
subset of the STAR*D data. This may be due to the lack of
heteroskedasticity at the first stage.  Thus, maximizing the
end-of-stage-two QIDS using mean-optimal techniques seems appropriate
and, in practice,   
equivalent to maximizing remission probabilities for each patient with 
TIQ-learning.   
 

\section{Discussion}
\label{sec:discussion}

We have proposed modeling frameworks for estimating
optimal dynamic treatment regimes in settings where a non-mean
distributional summary is the intended outcome to optimize.  Threshold
Interactive $Q$-learning estimates a regime that maximizes the mass of
the response distribution that exceeds a constant or patient-dependent
threshold, and Quantile Interactive $Q$-learning maximizes a
prespecified quantile of the response distribution.  

An important
application of TIQ-learning is to estimate a regime that maximizes
the probability of achieving remission for each patient, where
remission is defined in terms of an indicator of threshold exceedance.
Although we advocate for using the suggested interactive model
building tools, it is possible to prespecify a class of regimes and
directly optimize within that class by utilizing the value function, 
\begin{eqnarray*}
V_{\hbox{\scriptsize TIQ}}^{\hbox{\scriptsize $\pi_{1}$},
  \hbox{\scriptsize $\pi_{2}$}} =
\frac{\sum_{i=1}^{n} \mathbbm{1}_{Y_{i} > \lambda} \mathbbm{1}_{A_{1, i} =
    \pi_{1}(\hbox{\scriptsize $\bH_{1, i}$} )}\mathbbm{1}_{A_{2, i} =
    \pi_{2}(\hbox{\scriptsize $\bH_{2, i}$} ) } }{\sum_{i=1}^{n}\mathbbm{1}_{A_{1, i} =
    \pi_{1}(\hbox{\scriptsize $\bH_{1, i}$} )}\mathbbm{1}_{A_{2, i} =
    \pi_{2}(\hbox{\scriptsize $\bH_{2, i}$} ) } },  
\end{eqnarray*}
for TIQ-learning. In addition, it has been shown
  that $Q$-learning, $A$-learning, and $g$-estimation lead to
  identical estimators in certain cases but that $Q$-learning is less
  efficient \citep{Cha+etal:10, Sch+etal:12}. It
     is possible a g-estimation approach exists for maximizing
     probabilities and quantiles based on structural nested
     distribution models \citep{rob:00, van+jof:14}. This is an open
     question. 

Our proposed
methods are designed for the two-stage setting, this is
an important development given that many completed and ongoing SMART  
studies have this structure \citep{psu:12, eblSmart}. Here we
considered binary treatments at both stages.  In 
principle, the proposed methods can be extended to settings with more
than two treatments at each stage by modeling additional
treatment contrasts. Formalization of this idea merits further research.

\section{Supplementary Materials}
\label{sec:supplement}

Online supplementary materials include 
discussions of modeling adjustments for heteroskedastic
second-stage errors and patient-specific thresholds, a proof of Lemma
\ref{lemma1} and toy example illustrating where this
lemma does not apply, additional simulation results, and proofs of the
theorems in Section \ref{subsec:theory}.

\baselineskip13pt

\bibliographystyle{apalike}
\bibliography{iq_cites}
  
\end{document}